\documentclass[letterpaper]{article} 
\usepackage{aaai2027}  

\usepackage[hyphens]{url}  
\usepackage{graphicx} 
\usepackage{natbib}  
\usepackage{caption} 
\usepackage{algorithm}

\usepackage{algpseudocode}

\usepackage{newfloat}
\usepackage{listings}
\usepackage{amsfonts}

\usepackage{enumitem}
\usepackage{multirow}
\usepackage{amsmath}    
\usepackage{extarrows}  
\usepackage[table]{xcolor}
\usepackage{subcaption}
\usepackage{bm}

\DeclareCaptionStyle{ruled}{labelfont=normalfont,labelsep=colon,strut=off} 
\floatstyle{ruled}
\newfloat{listing}{tb}{lst}{}
\floatname{listing}{Listing}

\usepackage{booktabs}

\newcommand{\model}{\textsl{PrismRec} }
\newcommand{\modelwospace}{\textsl{PrismRec}}

\title{Preference Flow Matching with Spectral Factorization for Micro-video Recommendation}

\author{
    Xinxin Dong\textsuperscript{\rm 1},
    Haokai Ma\textsuperscript{\rm 2}\thanks{Corresponding Author.},
    Fei Hu\textsuperscript{\rm 1},\\
    YuZe Zheng\textsuperscript{\rm 1},
    Bin Wu\textsuperscript{\rm 3},
    Yonghui Yang\textsuperscript{\rm 2},
    Xiaodong Wang\textsuperscript{\rm 1},
}
\affiliations{
    \textsuperscript{\rm 1}National Key Laboratory of Parallel and Distributed Computing, College of Computer Science and Technology, \\ National University of Defense Technology\\

    \textsuperscript{\rm 2}National University of Singapore\\
    \textsuperscript{\rm 3}Zhengzhou University\\

haokai.ma@nus.edu.sg
}

\nocopyright
\begin{document}

\maketitle

\begin{abstract}
Micro-video recommendation aims to infer user preferences from historical interactions and multimodal video content, thereby identifying the next video of interest. However, prevailing methods compress frame sequences into a single holistic representation, entangling the stable visual semantics and the evolving dynamics that jointly shape user preferences. Meanwhile, diffusion- and flow matching-based recommenders condition their generation process solely on coarse behavioral context, leaving its internal temporal structure outside preference formation. 
We therefore propose \textit{PrismRec}, a \textbf{Pr}eference Flow Match\textbf{i}ng framework with \textbf{S}pectral Factorization for \textbf{M}icro-video \textbf{Rec}ommendation. 
Analogous to a prism that disperses white light into its constituent spectrum, \textit{PrismRec} devises Spectral Semantic Factorization (SSF) to derive complementary static semantic and dynamic factors from frame-level representations via a prior-guided learnable frequency mask in the temporal frequency domain. 
Then, it proposes Context-Calibrated Preference Matching (CPM) to weigh them with each user's specific sensitivity and inject the calibrated context as a structured condition to steer the matching trajectory toward the target representation, making video content as an intrinsic driver of preference formation rather than auxiliary side information. 
Experiments on four datasets from two platforms show that \textit{PrismRec} surpasses the SOTA baseline by up to 22.65\%, with the lowest inference cost and peak memory among the compared methods.
\end{abstract}

\section{Introduction}

With the rapid proliferation of video platforms, micro-videos have emerged as an increasingly important form of content consumption~\cite{MicroLens}. Micro-video recommendation aims to infer user preferences from both the historical interactions and the multi-modal content of interacted videos, thereby identifying the next video they are likely to consume~\cite{cai2024popularity, zhao2019recommending,he2025user}. Compared with conventional recommendation, the content herein carries richer preference evidence, as stable visual semantics and evolving frame-level dynamics jointly shape user interests, making the exploitation of such temporally structured content central to accurate preference modeling.

\begin{figure}[t]
    \centering
    \includegraphics[width=1.0\linewidth]{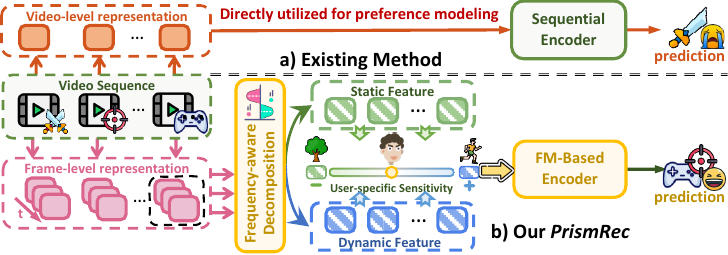}
    \caption{Motivation of our \textit{PrismRec}. Existing methods directly use holistic video-level representations for preference modeling, potentially obscuring temporally distinct content cues. \textit{PrismRec} factorizes frame-level representations into complementary static and dynamic factors and performs user-specific calibration to guide the preference matching trajectory toward fine-grained next-video prediction.}
    \label{fig:motivation_graph}
\end{figure}

Existing research on micro-video recommendation has largely evolved along two complementary lines that respectively focus on interaction-centric preference modeling and content-centric video representation. The former infers user interests from collaborative relations in user--video graphs~\cite{wei2019mmgcn} or sequential dependencies across users’ chronological viewing sequences~\cite{pan2023understanding}. However, it primarily captures statistical regularities in observed interactions, leaving the underlying process of preference formation unexplored. The latter enhances video understanding through adaptive multimodal fusion~\cite{liu2023dynamic, fu2025efficient} and segment-level interest modeling~\cite{shang2023learning,he2025short}. Nevertheless, these methods ultimately compress the extracted content cues into unified video representations, obscuring the distinction between static semantics and dynamic patterns. Consequently, current video content largely serves as auxiliary side information rather than the intrinsic component of preference dynamics.

Recent advances in computer vision offer two complementary tools to remedy these issues. On the content side, spectral analysis characterizes temporal patterns at different variation rates, where low-frequency components tend to encode stable visual semantics and static cues, whereas higher-frequency components are more associated with temporal variations, including motion patterns and transition dynamics~\cite{kim2023frequency,ponbagavathi2026frame2freq}. On the preference side, flow matching~\cite{lipman2023flow} (FM) formulates distribution matching as a continuous transport process by learning a vector field from a simple prior to the target distribution, and has recently been introduced into recommendation to characterize the continuous transition from noisy latent states to target item representations~\cite{liu2025fmrec, li2025flowrec}. However, these two tools have thus far evolved in isolation: spectral analysis remains confined to general video understanding without connecting the factorized structure to user preferences, while FM-based recommenders remain conditioned on interaction-centric signals, leaving the internal temporal structure of video content outside the generative process. This raises a central question: \emph{Can frequency-aware video factorization be integrated with continuous preference generation to make structured video content an intrinsic driving force of preference formation?}

Despite its promise, this paradigm poses dual challenges, that is, \textbf{temporally entangled content factors} and \textbf{content-agnostic preference generation}. Specifically, static semantics and dynamic variations coexist across the frame sequence at distinct temporal rates, yet prevailing pipelines aggregate them into a single holistic video representation, entangling the two signals during preference modeling (\emph{cf.} Figure~\ref{fig:motivation_graph} (a)). Nevertheless, granting such factors distinct roles is non-trivial for existing FM-based recommenders, whose generative processes are conditioned on coarse behavioral context. Naively injecting content features into the condition would collapse the factorized structure back into an undifferentiated signal, leaving the complementary guidance of static and dynamic factors unexploited along the matching trajectory.

To address these challenges, we propose a \textbf{Pr}eference Flow Match\textbf{i}ng framework with \textbf{S}pectral Factorization for \textbf{M}icro-video \textbf{Rec}ommendation, termed as \textsc{PrismRec}. Analogous to a prism that disperses white light into its constituent spectrum, \textsc{PrismRec} factorizes entangled frame-level representations into complementary static semantic and dynamic factors, and injects them as structured conditions to steer the matching trajectory of target preferences (\emph{cf.} Figure~\ref{fig:motivation_graph} (b)). Specifically, we first conduct empirical analysis in Section~\ref{sec:empirical_analysis} to reveal that the frequency composition varies substantially across videos and users exhibit consistent yet distinct sensitivities to static and dynamic content, indicating that the two factors deserve separate characterization. Motivated by this, we devise \textbf{Spectral Semantic Factorization (SSF)}, which transforms frame-level representations into the temporal frequency domain and derives complementary static semantic and dynamic factors via a learnable frequency mask with branch-specific asymmetric refinement. To further grant these factors distinct roles in preference formation, we design \textbf{Context-Calibrated Preference Matching (CPM)}, which formulates next-item prediction as a content-conditioned generation process. Here, the factorized factors in SSF and textual semantics serve as structured contextual conditions, jointly calibrating the intermediate preference states along the matching trajectory derived from collaborative signals and sequential dependencies. Equipped with such modules, structured video content acts as an intrinsic driving force of preference formation rather than the auxiliary side information.

Extensive experiments on four real-world micro-video datasets from two platforms demonstrat that \textsc{PrismRec} consistently outperforms representative baselines while maintaining strong robustness and efficiency. The main contributions are summarized as follows:
\begin{itemize}[leftmargin=*, topsep=0.2pt, parsep=0pt]
    \item We revisit micro-video recommendation from a unified perspective that couples frequency-aware content factorization with continuous preference generation. To the best of our knowledge, we are among the pioneer to formulate video content as an intrinsic driving force of the preference matching trajectory.
    \item We propose \textsc{PrismRec}, where SSF derives complementary static semantic and dynamic factors in the temporal frequency domain to resolve the entangled content factors, and CPM injects them as structured conditions to calibrate the content-agnostic matching trajectory for controllable preference generation.
    \item Extensive experiments on four real-world datasets, together with ablation study, robustness analysis and visualization analysis jointly indicate the superiority of our \textsc{PrismRec}.
\end{itemize}

\section{Related Work}

\paragraph{Micro-video Recommendation}
Recent micro-video recommendation has shifted from ID-based collaborative filtering to joint modeling of video content and fine-grained behavioral signals. Content-driven methods initially exploit multimodal side information to alleviate interaction sparsity~\cite{UVCAN,wei2019mmgcn}, while subsequent studies learn richer content semantics. MicroLens~\cite{MicroLens} enables representation learning from raw multimodal content, whereas FRAME~\cite{shang2023learning} captures local semantics and user interests through user–clip relations. However, these methods typically model videos as static features or isolated clips, without explicitly distinguishing static semantics from temporal dynamics. Behavior-driven methods extend sequential recommendation by modeling richer user behaviors and preference modeling strategies~\cite{YouTubeDNN,TriCDR,CPRec}. D$^2$Co~\cite{D2Co} mitigates watch-time bias, while fine-grained skip modeling~\cite{FineGrainedSkip} captures nuanced preference dynamics. Nevertheless, they primarily infer interests from interaction logs, leaving the temporal semantics of video content underexplored.

\paragraph{Frequency-Domain and Flow-Based Learning in Recommendation}
In computer vision, spectral analysis distinguishes low-frequency components associated with stable visual semantics from high-frequency components reflecting motion and transition dynamics~\cite{kim2023frequency,ponbagavathi2026frame2freq}. Inspired by this property, frequency-domain recommenders project interaction signals into spectral space for noise suppression and multi-scale preference modeling~\cite{zhou2022fmlprec,du2023fearec,cai2026dynamic}, with recent extensions incorporating multimodal information through spectral fusion and frequency-aware representation learning~\cite{ong2025smore,yang2026fitmm}. In parallel, diffusion-based recommenders model complex preference distributions via iterative denoising~\cite{wang2023diffrec,li2023diffurec,wang2024cddrec,PDRec}, but provide limited control over noise injection. Flow matching instead learns continuous transport trajectories from a prior to the target distribution~\cite{lipman2023flow,liu2023rectified} and has recently been adopted in various recommendation settings~\cite{liu2025flowcf,liu2025fmrec,shi2026fave}. Nevertheless, integrating frame-level spectral representations with flow-based preference generation remains underexplored.

\section{Preliminaries and Empirical Analysis}
\label{sec:preliminary}

\paragraph{Flow Matching}
\label{sec:flow_matching}
Flow Matching (FM) learns a continuous normalizing flow by regressing a time-dependent vector field along a prescribed probability path, thereby avoiding the simulation of a stochastic diffusion process~\cite{lipman2023flow}. Let $p_0$ and $p_1$ denote the source and target distributions, respectively. Conditional Flow Matching (CFM) constructs a tractable path between samples $\boldsymbol{x}_0 \sim p_0$ and $\boldsymbol{x}_1 \sim p_1$. For $t \in [0,1]$, we consider the affine path $\boldsymbol{x}_t=\alpha_t \boldsymbol{x}_1+\sigma_t \boldsymbol{x}_0$,
where $\alpha_t$ and $\sigma_t$ are differentiable schedules satisfying $\alpha_0=0$, $\sigma_0=1$, $\alpha_1=1$, and $\sigma_1=0$.
Differentiating this path yields the conditional velocity field
\begin{equation}
\boldsymbol{u}_t(\boldsymbol{x}_t \mid \boldsymbol{x}_1)=
\dot{\alpha}_t \boldsymbol{x}_1+\dot{\sigma}_t\frac{\boldsymbol{x}_t-\alpha_t\boldsymbol{x}_1}{\sigma_t}
\end{equation}
where $\dot{\alpha}_t$ and $\dot{\sigma}_t$ denote the corresponding time derivatives. A neural vector field $\boldsymbol{v}_{\theta}(\boldsymbol{x}_t,t)$ is then optimized to approximate this conditional velocity through
\begingroup
\small
\begin{equation}
\mathcal{L}_{\mathrm{CFM}}(\theta)=\mathbb{E}_{\substack{t \sim \mathcal{U}(0,1),\,\boldsymbol{x}_0 \sim p_0,\,\boldsymbol{x}_1 \sim p_1}}\left[\left\|\boldsymbol{v}_{\theta}(\boldsymbol{x}_t,t) - \boldsymbol{u}_t(\boldsymbol{x}_t \mid \boldsymbol{x}_1)\right\|_2^2\right]
\end{equation}
\endgroup
At inference, samples are generated by integrating the learned ordinary differential equation (ODE)
\begin{equation}
\frac{d\boldsymbol{x}_t}{dt}=\boldsymbol{v}_{\theta}(\boldsymbol{x}_t,t),\qquad\boldsymbol{x}_{t=0} \sim p_0
\end{equation}
from $t=0$ to $t=1$, such that the terminal state $\boldsymbol{x}_{t=1}$ approximates a sample from $p_1$.

Compared with diffusion models, FM directly models continuous transport trajectories in representation space, avoiding iterative noising and denoising. This formulation naturally supports conditional modeling by allowing auxiliary signals to guide the entire transport process.

\begin{figure}[t]
    \centering
    \includegraphics[width=\linewidth]{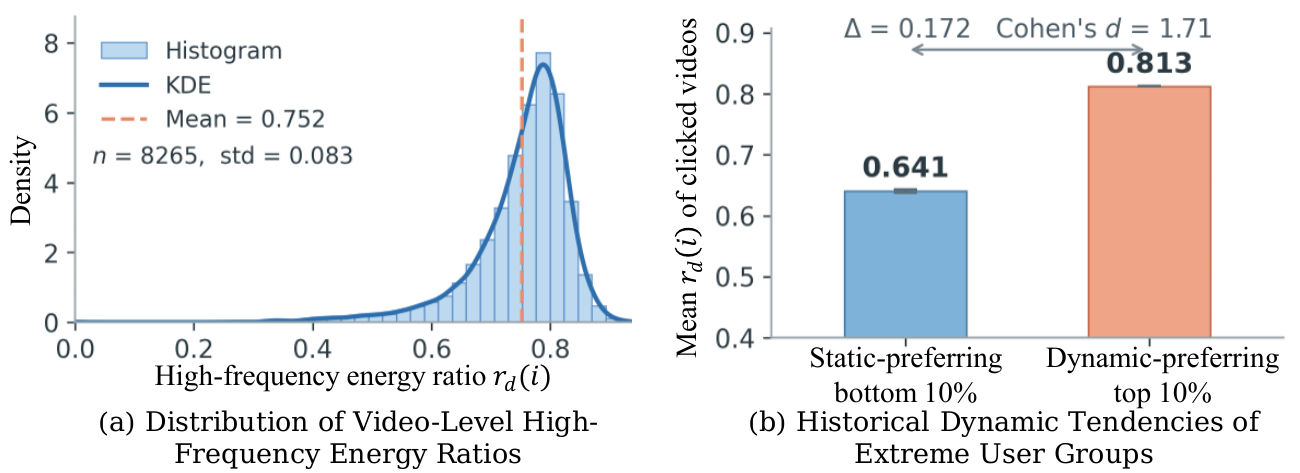}
    \caption{Empirical results of video-level temporal dynamics and user-level preference heterogeneity on MicroLens-Small.}
    \label{fig:empirical_analysis}
\end{figure}

\paragraph{Empirical Analysis}
\label{sec:empirical_analysis}

We conduct an empirical study on MicroLens-Small to examine the temporal characteristics of video representations and their relation to user preferences. Following frequency-domain analysis, we quantify the dynamic tendency of each video using the dynamic energy ratio:
\begin{equation}
r_d(i)=\frac{\sum_{k\in\mathcal{H}}\|\mathbf{X}_i(k)\|_2^2}{\sum_k\|\mathbf{X}_i(k)\|_2^2}
\end{equation}
where $\mathbf{X}_i(k)$ denotes the $k$-th frequency component of the frame-level representation of video $i$ after temporal FFT, and $\mathcal{H}$ represents the set of high-frequency components. A larger $r_d(i)$ indicates stronger temporal variations in the video. As shown in Fig.~\ref{fig:empirical_analysis}(a), $r_d(i)$ exhibits a highly non-uniform distribution, revealing substantial diversity in temporal patterns across videos. This indicates that videos contain varying degrees of static semantics and temporal dynamics, which may be overlooked by holistic content modeling. We further characterize user-level preferences by averaging $r_d(i)$ over the videos consumed in each user's training history. Fig.~\ref{fig:empirical_analysis}(b) reveals that users exhibit different tendencies toward static and dynamic video content. These findings suggest that static semantics and temporal dynamics capture complementary aspects of video preference, motivating their explicit modeling in recommendation.

\begin{figure*}[t]
    \includegraphics[width=0.9\linewidth]{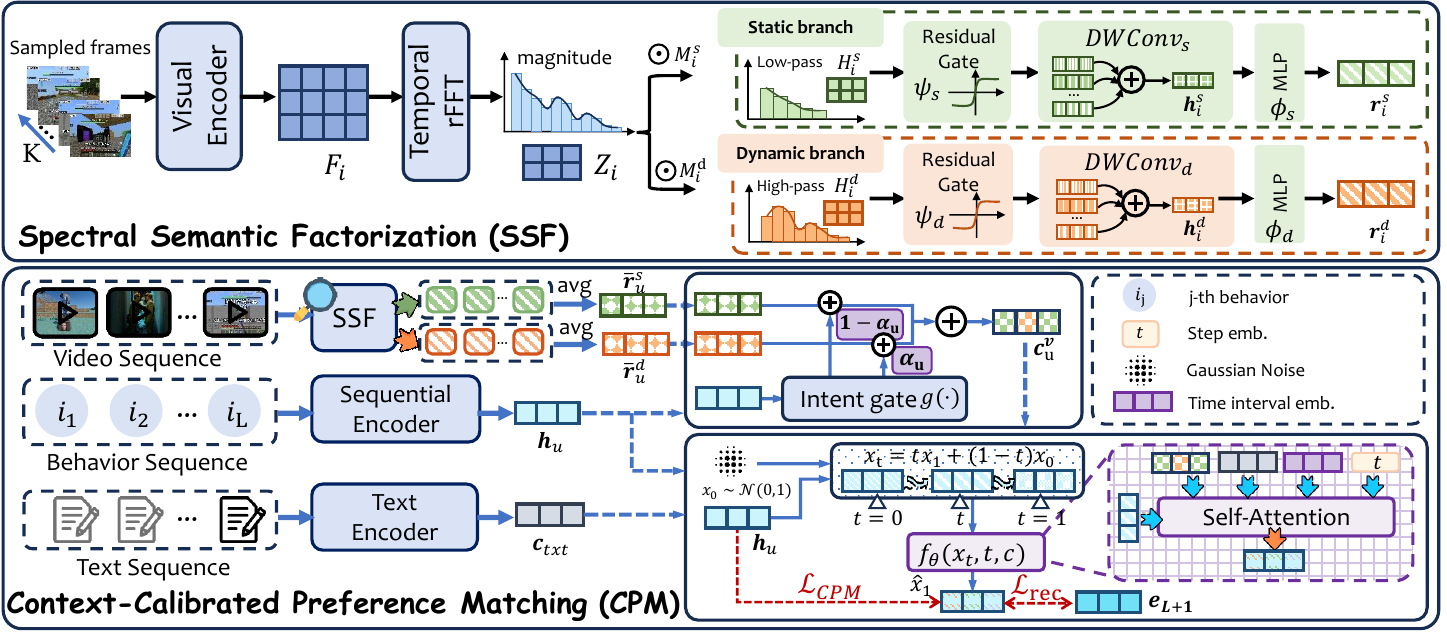}
    \centering
    \caption{Overall structure of our proposed \modelwospace.}
    \label{fig:overall_stracture}
    \vspace{-0.3cm}
\end{figure*}

\section{Methodology}
\label{sec:method}

We present \modelwospace, a multimodal flow matching framework for micro-video sequential recommendation, consisting of two key components. SSF decomposes frame-level video representations into complementary static and dynamic factors through temporal spectral analysis. CPM models personalized preference evolution via conditional flow matching, where the disentangled video factors and multimodal contexts jointly guide trajectory learning.

\subsection{Problem Formulation}
\label{sec:problem}
Let $\mathcal{U}$ and $\mathcal{I}$ denote the sets of users and items, respectively. For each user $u\in\mathcal{U}$, let $\mathcal{S}_u=[(i_1,\tau_1),\ldots,(i_L,\tau_L)]$
denote the chronological interaction sequence, where $i_l\in\mathcal{I}$ is the video consumed at timestamp $\tau_l$. Micro-video sequential recommendation aims to predict the next video $i_{L+1}$ given $\mathcal{S}_u$. Each video $i$ is associated with multimodal content, including frame-level visual features
$\mathbf{F}_i=[\mathbf{f}_{i,1},\ldots,\mathbf{f}_{i,K}]\in\mathbb{R}^{K\times d}$
and a textual representation $\mathbf{E}^{\mathrm{txt}}_i$, where $K$ is the number of sampled frames and $d_{vis}$ is the visual feature dimension.

\subsection{Spectral Semantic Factorization (SSF)}
\label{sec:ssf}
Micro-videos simultaneously contain relatively stable visual semantics and rapidly varying temporal patterns. Directly compressing frame-level features into a single holistic representation may obscure these heterogeneous signals. We therefore propose SSF, which consists of prior-guided spectral decomposition, branch-specific frequency refinement, and asymmetric dual-track fusion.

Given the frame-level representation $\mathbf{F}_i\in\mathbb{R}^{K\times d}$, we first apply the real FFT along the temporal dimension to obtain the magnitude spectrum $\mathbf{Z}_i=|\operatorname{rFFT}_{t}(\mathbf{F}_i)|$. To separate stable visual semantics from rapidly varying temporal patterns, we introduce a prior-guided adaptive decomposition module. Specifically, a predefined low-/high-frequency prior $\mathbf{P}$ is combined with a video-specific residual $G(\mathbf{Z}_i)$ to generate two complementary soft masks:
\begin{equation}
\mathbf{M}_i^s=\sigma\!\left(\mathbf{P}+\eta G(\mathbf{Z}_i)\right),\qquad
\mathbf{M}_i^d=\mathbf{1}-\mathbf{M}_i^s
\label{eq:adaptive_masks}
\end{equation}
where $\eta$ controls the magnitude of the adaptive correction. The decomposition is initialized according to the predefined spectral prior and progressively adapts to individual videos during training. The resulting static- and dynamic-oriented spectral factors are $\mathbf{H}_i^s=\mathbf{M}_i^s\odot\mathbf{Z}_i$ and $\mathbf{H}_i^d=\mathbf{M}_i^d\odot\mathbf{Z}_i$, where $\odot$ denotes element-wise multiplication.

The two spectral factors are subsequently processed by branch-specific refinement modules. For each branch $b\in\{s,d\}$, we first recalibrate its channel responses using a residual gate, yielding $\overline{\mathbf{H}}_i^b=\mathbf{H}_i^b\odot(\mathbf{1}+\gamma\mathbf{a}_i^b)$, where $\mathbf{a}_i^b=\tanh(\psi_b(\operatorname{Avg}_{f}(\mathbf{H}_i^b)))$ and $\gamma$ controls the residual adjustment. We then employ branch-specific multi-scale depth-wise convolutions to capture local dependencies among neighboring frequency bins:
\begin{equation}
\widetilde{\mathbf{H}}_i^b=
\operatorname{Proj}_b\left(
\mathop{\operatorname{Concat}}_{\kappa\in\mathcal{Q}_b}
\operatorname{GELU}\left(
\operatorname{DWConv}_{\kappa}^{b}(\overline{\mathbf{H}}_i^b)
\right)
\right)
\label{eq:frequency_adaptation}
\end{equation}
where $\mathcal{Q}_b$ denotes the convolutional kernel set of branch $b$. Frequency-wise average pooling then produces the branch-level representation $\mathbf{h}_i^b=\operatorname{Avg}_{f}(\widetilde{\mathbf{H}}_i^b)$.

Finally, SSF adopts an asymmetric dual-track fusion strategy. Let $\mathbf{g}_i$ denote the global video representation extracted by the visual encoder. Since $\mathbf{g}_i$ already retains substantial appearance information, the static factor serves as a residual correction, whereas the dynamic factor forms an independent temporal track:
\begin{equation}
\mathbf{r}_i^s=\mathbf{g}_i+\lambda_s\phi_s\!\left([\mathbf{h}_i^s;\mathbf{g}_i]\right),\qquad
\mathbf{r}_i^d=\phi_d(\mathbf{h}_i^d)
\label{eq:dual_track_fusion}
\end{equation}
where $[\cdot;\cdot]$ denotes concatenation and $\lambda_s$ controls the static residual strength. The resulting representations $\mathbf{r}_i^s$ and $\mathbf{r}_i^d$ provide complementary static- and dynamic-oriented content signals for subsequent preference modeling.

\subsection{Context-Calibrated Preference Matching (CPM)}
\label{sec:cpm}

Although SSF disentangles micro-video content into static and dynamic factors, their contributions to a user's next preference depend on behavioral, multimodal, and temporal contexts. Existing sequential recommenders typically compress interaction history into a single latent state, leaving its transition toward the next-item representation implicit. We therefore propose CPM, which models this transition as a context-calibrated transport process in latent space.

\paragraph{Context-Calibrated Preference Representation.}
Given the historical interaction sequence $\mathcal{S}_u$, we obtain the sequential preference state $\mathbf{h}_u=\operatorname{SeqEnc}(\mathcal{S}_u)$. For each historical video $i_l$, SSF produces the static and dynamic representations $\mathbf{r}_{i_l}^{s}$ and $\mathbf{r}_{i_l}^{d}$. We aggregate them over the interaction history as $\overline{\mathbf{r}}_u^b=\operatorname{Avg}_{i_l\in\mathcal{S}_u}(\mathbf{r}_{i_l}^b)$ for $b\in\{s,d\}$. Since users may exhibit different sensitivities to static semantics and temporal dynamics, CPM derives an intent gate from the sequential state and uses it to calibrate the two video factors:
\begin{equation}
\alpha_u=\sigma\!\left(g(\mathbf{h}_u)\right),\qquad
\mathbf{c}_u^v=(1-\alpha_u)\overline{\mathbf{r}}_u^s+\alpha_u\overline{\mathbf{r}}_u^d
\label{eq:video_context}
\end{equation}
where $\mathbf{c}_u^v$ denotes the calibrated video context. The textual context is obtained by averaging the textual representations of historical videos, denoted by $\mathbf{c}_u^{\mathrm{txt}}=\operatorname{Avg}_{i_l\in\mathcal{S}_u}(\mathbf{E}_{i_l}^{\mathrm{txt}})$. We further encode the interval between the historical sequence and the target interaction as $\mathbf{e}_u^\Delta=\psi_\Delta(\log(1+\Delta\tau_u/\tau_0))$, where $\tau_0$ is a scaling constant.

\paragraph{Context-Calibrated Preference Flow.}
Let $\mathbf{x}_1\!=\!\mathbf{e}_{i_{L+1}}$ denote the embedding of the next interacted item and $\mathbf{x}_0\!\sim\!\mathcal{N}(\mathbf{0},\mathbf{I})$ denote the initial state. Following a linear rectified-flow path, we construct $\mathbf{x}_t=(1-t)\mathbf{x}_0+t\mathbf{x}_1$ for $t\in[0,1]$. At each transport step, the current flow state, flow-time embedding, and temporal interval are injected into the historical preference:
\begin{equation}
\widetilde{\mathbf{h}}_{u,t} = \mathbf{h}_u + \boldsymbol{\xi}_t \odot \left( \mathbf{x}_t+\mathbf{e}_t \right) + \mathbf{e}_u^\Delta 
\end{equation}
where $\mathbf{e}_t$ is the flow-time embedding and $\boldsymbol{\xi}_t$ provides stochastic modulation of the evolving
state.
The trajectory-aware preference, calibrated video context, and textual context are jointly integrated to predict the target endpoint:
\begin{equation}
\widehat{\mathbf{x}}_1 = f_{\theta} \left( \widetilde{\mathbf{h}}_{u,t}, \mathbf{c}_u^{v}, \mathbf{c}_u^{\mathrm{txt}} \right)
\end{equation}
where $f_{\theta}(\cdot)$ denotes the context-calibrated prediction network. CPM thus adopts an $x_1$-parameterization, directly estimating the target rather than the velocity field.

At inference, the predicted endpoint induces the transport direction
$\mathbf{v}_{\theta}^{(n)} =\widehat{\mathbf{x}}_1^{(n)}-\mathbf{x}_0$.
Starting from $\mathbf{x}^{(0)}=\mathbf{x}_0$, CPM performs $N$ Euler updates:
\begin{equation}
\mathbf{x}^{(n+1)} \!=\!\mathbf{x}^{(n)} \!+\!\frac{1}{N} \left(\widehat{\mathbf{x}}_1^{(n)}\!-\!\mathbf{x}_0\right),
\quad
n\!=\!0,\ldots,N-1
\label{eq:euler_transport}
\end{equation}
The terminal representation $\mathbf{x}_u^{\mathrm{FM}}\!=\!\mathbf{x}^{(N)}$ is finally combined with the sequential preference representation as $\mathbf{p}_u\!=\! \omega\mathbf{x}_u^{\mathrm{FM}} \!+\!(1\!-\!\omega)\mathbf{h}_u$, preserving the collaborative signals encoded in the interaction history.

\subsection{Optimization Objective}
\label{sec:optimization}

We optimize the entire framework end-to-end with two complementary objectives. Under the $x_1$-parameterization, the flow-matching objective aligns the predicted endpoint with the representation of the observed next item:
\begin{equation}
\mathcal{L}_{\mathrm{CPM}} = \mathbb{E}_{t,\bm{x}_0,\bm{x}_1} \left[ \left\|\hat{\bm{x}}_1-\bm{x}_1\right\|_2^2\right]\label{eq:fm_loss}
\end{equation}
This representation-level supervision guides the context-conditioned transport process toward the target preference state. To ensure that the resulting user representation remains discriminative over the item space~\cite{NS4RS}, we further employ a full-softmax recommendation objective:
\begin{equation}
\mathcal{L}_{\mathrm{rec}}=-\frac{1}{|\mathcal{B}|}\sum_{u\in\mathcal{B}}\log\frac{\exp\!\left(s(u,i_u^+)\right)}{\sum_{j\in\mathcal{I}}\exp\!\left(s(u,j)\right)}
\label{eq:rec_loss}
\end{equation}
where $\mathcal{B}$ denotes a training batch and $i_u^+$ is the observed next item of user $u$. While $\mathcal{L}_{\mathrm{FM}}$ regularizes preference evolution in the representation space, $\mathcal{L}_{\mathrm{rec}}$ directly optimizes next-item discrimination. The overall objective is defined as
\begin{equation}
\mathcal{L} =\mathcal{L}_{\mathrm{rec}}+\lambda_{\mathrm{C}}\mathcal{L}_{\mathrm{CPM}}
\label{eq:overall_loss}
\end{equation}
where $\lambda_{\mathrm{C}}$ controls the contribution of $\mathcal{L}_{\mathrm{CPM}}$. We futher conduct complexity analysis of our method in Appendix.

\section{Experiments}
\label{sec:experiments}

We conduct extensive experiments to address the following research questions:

\noindent
\textbf{RQ1.} How does \model compare with fourteen methods on four real-world datasets?

\noindent
\textbf{RQ2.} How does each key component of \model contribute to the overall recommendation performance?

\noindent
\textbf{RQ3.} How does \model generalize across different sequential recommendation backbones?

\noindent
\textbf{RQ4.} How sensitive is \model to the key hyperparameters governing spectral factorization and preference transport?

\noindent
\textbf{RQ5.} How robust and efficient is \model under practical variations in input noise and frame sampling?

\subsection{Experimental Setups}

\subsubsection{Datasets.}
We conduct experiments on four real-world micro-video recommendation benchmarks collected from two platforms, including MicroLens-Small and MicroLens-Big~\cite{MicroLens}, as well as Shortvideo-Small and Shortvideo-Big~\cite{shang2025large}. 
These benchmarks provide rich multimodal video information, including raw videos, textual metadata, and associated user-video interaction records with timestamps, enabling comprehensive modeling of video semantics and user preference evolution. Following established preprocessing protocols, we apply a 4-core filtering strategy and organize user interactions into chronological sequences. We adopt the leave-one-out evaluation protocol, where the latest and second latest interactions are held out for testing and validation, respectively. The detailed dataset statistics are summarized in Table~\ref{tab:dataset_stats}.

\begin{table}[t]
\centering
\caption{The detailed statistics of four real-world datasets.}
\label{tab:dataset_stats}
\resizebox{\columnwidth}{!}{%
\begin{tabular}{lcccc}
\toprule
Dataset & \#Users & \#Items & \#Interactions & Sparsity \\
\midrule
Microlens-Small  & 8,694  & 8,265   & 50,838   & 99.93\% \\
Microlens-Big    & 86,073 & 18,700  & 597,104  & 99.96\% \\
Shortvideo-Small & 3,460  & 84,192  & 240,843  & 99.92\% \\
Shortvideo-Big   & 7,584  & 122,845 & 533,381  & 99.94\% \\
\bottomrule
\end{tabular}}
\end{table}

\subsubsection{Baselines}
We compare \model with fourteen baselines from four categories: collaborative filtering methods (i.e., \textbf{BPR}, \textbf{VBPR}, and \textbf{FlowCF}), sequential recommenders (i.e., \textbf{SASRec}, 
\textbf{CL4SRec}, \textbf{SSDRec}, \textbf{DiQDiff}, and \textbf{FMRec}), multimodal sequential recommenders (i.e., \textbf{MoRec}, \textbf{TedRec}, \textbf{IISAN}, and \textbf{DMMD4SR}), and video recommenders 
(i.e., \textbf{MMGCN} and \textbf{IISAN-Versa}). Detailed descriptions of these baselines are provided in the
Appendix.

\begin{table*}[t]
\centering
\caption{Performance comparison of our \model against fourteen baselines across four micro-video datasets from two platforms. $^{*}$ denotes significant improvements of \model over the baselines (\emph{p} $\textless$ 0.01 with paired t-tests).}
\label{tab:performance_comparison_extended}
\setlength{\tabcolsep}{1.7pt}
\resizebox{\textwidth}{!}{
\begin{tabular}{lcccccccccccccccc}
\toprule
\multirow{2}{*}{\textbf{Algorithm}} & \multicolumn{4}{c}{\textbf{Microlens-Small}} & \multicolumn{4}{c}{\textbf{Microlens-Big}} & \multicolumn{4}{c}{\textbf{Shortvideo-Small}}  & \multicolumn{4}{c}{\textbf{Shortvideo-Big}} \\
\cmidrule(lr){2-5} \cmidrule(lr){6-9} \cmidrule(lr){10-13} \cmidrule(lr){14-17}
 & \textbf{H@10} & \textbf{N@10} & \textbf{H@20} & \textbf{N@20} & \textbf{H@10} & \textbf{N@10} & \textbf{H@20} & \textbf{N@20}  
 & \textbf{H@10} & \textbf{N@10} & \textbf{H@20} & \textbf{N@20} & \textbf{H@10} & \textbf{N@10} & \textbf{H@20} & \textbf{N@20} \\
\midrule

BPR (UAI '09)       & 0.0177 & 0.0064 & 0.0613 & 0.0171 & 0.0085 & 0.0040 & 0.0157 & 0.0058 & 0.0116 & 0.0061 & 0.0225 & 0.0088 & 0.0113 & 0.0057 & 0.0183 & 0.0074 \\
VBPR (AAAI '16)     & 0.0202 & 0.0069 & 0.0648 & 0.0179 & 0.0128 & 0.0050 & 0.0294 & 0.0091 & 0.0133 & 0.0072 & 0.0231 & 0.0097 & 0.0125 & 0.0060 & 0.0219 & 0.0084 \\
FlowCF (KDD '25)  & 0.0261 & 0.0103 & 0.0719 & 0.0217 & 0.0147 & 0.0064 & 0.0318 & 0.0107 & 0.0147 & 0.0079 & 0.0228 & 0.0099 & 0.0166 & 0.0088 & 0.0286 & 0.0119 \\
\midrule
SASRec (ICDM '18)   & 0.0741 & 0.0414 & 0.0932 & 0.0462 & 0.0736 & 0.0366 & 0.1115 & 0.0462 & 0.0150 & 0.0069 & 0.0237 & 0.0091 & 0.0185 & 0.0092 & 0.0306 & 0.0123 \\  
CL4SRec (ICDE '22)  & 0.0764 & 0.0419 & 0.0941 & 0.0464 & 0.0748 & 0.0384 & 0.1127 & 0.0479 & 0.0162 & 0.0083 & 0.0257 & 0.0107 & 0.0200 & 0.0098 & 0.0301 & 0.0123 \\
SSDRec (ICDE '24)   & 0.0790 & 0.0430 & 0.0975 & 0.0478 & 0.0624 & 0.0295 & 0.0942 & 0.0374 & 0.0121 & 0.0069 & 0.0153 & 0.0077 & 0.0153 & 0.0084 & 0.0199 & 0.0096 \\
DiQDiff (WWW '25)   & 0.0614 & 0.0415 & 0.0698 & 0.0437 & 0.0788 & 0.0430 & 0.1169 & 0.0526 & 0.0131 & 0.0060 & 0.0191 & 0.0075 & 0.0218 & 0.0105 & 0.0318 & 0.0130 \\
FMRec (IJCAI '25)     & 0.0619 & 0.0414 & 0.0715 & 0.0438 & \underline{0.0825} & \underline{0.0435} & \underline{0.1235} & \underline{0.0538} & 0.0180 & \underline{0.0106} & 0.0279 & \underline{0.0131} & \underline{0.0253} & \underline{0.0138} & \underline{0.0402} & \underline{0.0175} \\
\midrule
   
MoRec (SIGIR '23)   & 0.0786 & 0.0381 & 0.1064 & 0.0452 & 0.0615 & 0.0313 & 0.0962 & 0.0400 & 0.0122 & 0.0062 & 0.0187 & 0.0078 & 0.0125 & 0.0063 & 0.0173 & 0.0075 \\
TedRec (CIKM '24)   & 0.0798 & 0.0445 & 0.1055 & 0.0510 & 0.0760 & 0.0397 & 0.1155 & 0.0496 & \underline{0.0194} & 0.0092 & \underline{0.0309} & 0.0121 & 0.0237 & 0.0113 & 0.0357 & 0.0143 \\     
IISAN (SIGIR '24)   & 0.0810 & 0.0365 & 0.1048 & 0.0427 & 0.0767 & 0.0399 & 0.1165 & 0.0499 & 0.0159 & 0.0081 & 0.0225 & 0.0098 & 0.0118 & 0.0059 & 0.0157 & 0.0069 \\
DMMD4SR (MM '25)    & \underline{0.0842} & \underline{0.0461} & 0.1050 & 0.0514 & 0.0801 & 0.0383 & 0.1171 & 0.0476 & 0.0145 & 0.0067 & 0.0231 & 0.0089 & 0.0187 & 0.0099 & 0.0303 & 0.0129 \\ 
\midrule     

MMGCN (MM '19)      & 0.0349 & 0.0137 & 0.0779 & 0.0247 & 0.0202 & 0.0092 & 0.0372 & 0.0135 & 0.0156 & 0.0062 & 0.0182 & 0.0080 & 0.0131 & 0.0066 & 0.0204 & 0.0085 \\
IISAN-Versa (TKDE '25) & 0.0836 & 0.0454 & \underline{0.1076} & \underline{0.0516} & 0.0780 & 0.0422 & 0.1146 & 0.0514 & 0.0145 & 0.0069 & 0.0202 & 0.0083 & 0.0219 & 0.0116 & 0.0332 & 0.0144 \\
\midrule

 \textbf{\model (Ours)} & \textbf{0.0902*} & \textbf{0.0478*} & \textbf{0.1177*} & \textbf{0.0548*} & \textbf{0.0867*} & \textbf{0.0458*} & \textbf{0.1289*} & \textbf{0.0564*} & \textbf{0.0231*} & \textbf{0.0124*} & \textbf{0.0379*} & \textbf{0.0160*} & \textbf{0.0302*} & \textbf{0.0148*} & \textbf{0.0444*} & \textbf{0.0184*} \\
\textit{Rel. Imp.} & \textit{7.13\%} & \textit{3.69\%} & \textit{9.39\%} & \textit{6.20\%} & \textit{5.09\%} & \textit{5.29\%} & \textit{4.37\%} & \textit{4.83\%} & \textit{19.07\%} & \textit{16.98\%} & \textit{22.65\%} & \textit{22.14\%} & \textit{19.37\%} & \textit{7.25\%} & \textit{10.45\%} & \textit{5.14\%} \\ 
\bottomrule
\end{tabular}
}
\end{table*}

\begin{figure*}[t]
    \centering
    \includegraphics[width=1\linewidth]{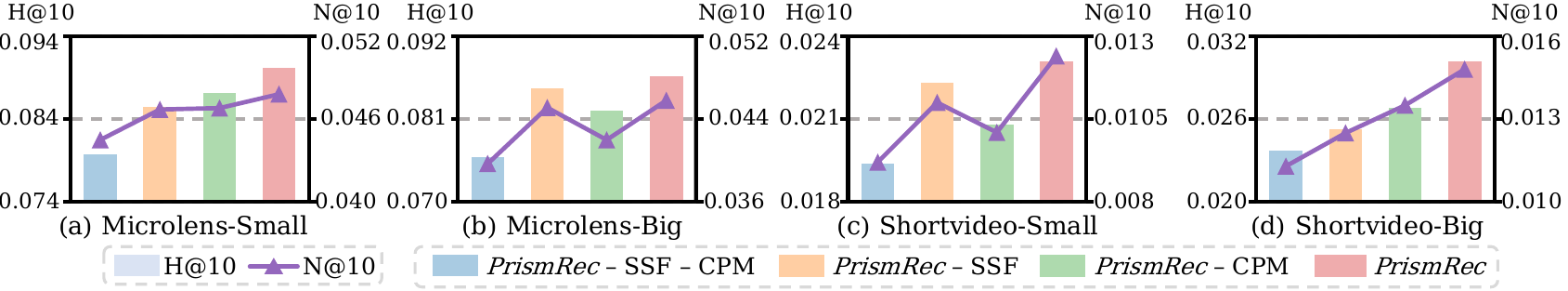}
    \vspace{-0.5cm}
    \caption{Ablation results on four benchmark datasets. Both components within our \textit{PrismRec} are effective.}
    \vspace{-0.5cm}
    \label{fig:prismrec_ablation}
\end{figure*}

\paragraph{Implementation Details.}
All experiments are conducted on NVIDIA 4090 GPU using Python 3.10.18.  We extract frame-level visual features with VideoMAE~\cite{tong2022videomae} from 10 uniformly sampled frames per video, and use BERT~\cite{devlin2019bert} and GloVe~\cite{pennington2014glove} for textual features on MicroLens and Short-video, respectively. The default sequential encoder contains two Transformer layers with 64 hidden dimensions, two attention heads, a feed-forward dimension of 2048, and a dropout rate of 0.5, while user sequences are truncated to 50 interactions.
All models are optimized with Adam using a batch size of 2048 and a learning rate of $10^{-3}$. SSF adopts orthonormal FFT, while CPM uses 10 sampling steps. The fusion weight $w$ and CPM loss weight $\lambda_{\mathrm{C}}$ are selected from $\{0.1,0.2,0.3,0.4,0.5\}$ based on the validation set.

\subsection{Overall Performance (RQ1)}
Table~\ref{tab:performance_comparison_extended} reports the performance comparison between \model and fourteen baselines across four datasets, from which we can observe that: 
(1) The margin of sequential recommenders over collaborative filtering is dataset-dependent, reaching 3.03 and 5.61 times in H@10 on MicroLens but only 1.22 and 1.52 times on ShortVideo, where a much larger item catalog leaves fewer than five interactions per item. The bottleneck therefore shifts from behavioral signal to item representation. 
(2) Multimodal content is not uniformly beneficial, since MoRec, IISAN, and DMMD4SR all fall below the purely sequential FMRec on ShortVideo datasets, suggesting that unified fusion strategies on video content inherits item-side sparsity rather than alleviating it. 
(3) FMRec attains the best baseline results on ten of the sixteen dataset-metric pairs, confirming that flow matching paradigm transport is well suited to preference modeling. The fact that our \model significantly improves over it indicating that our gains also stem from the structured content condition rather than solely from flow matching itself. 
(4) \model achieves the best performance on all sixteen dataset-metric pairs with $p\!<\!0.01$ and the gain is most pronounced exactly where interaction-derived item embeddings are least reliable. This is consistent with treating factorized video content as an intrinsic driving force of preference formation rather than auxiliary side information.

\begin{figure}[t]
    \centering
    \includegraphics[width=1\linewidth]{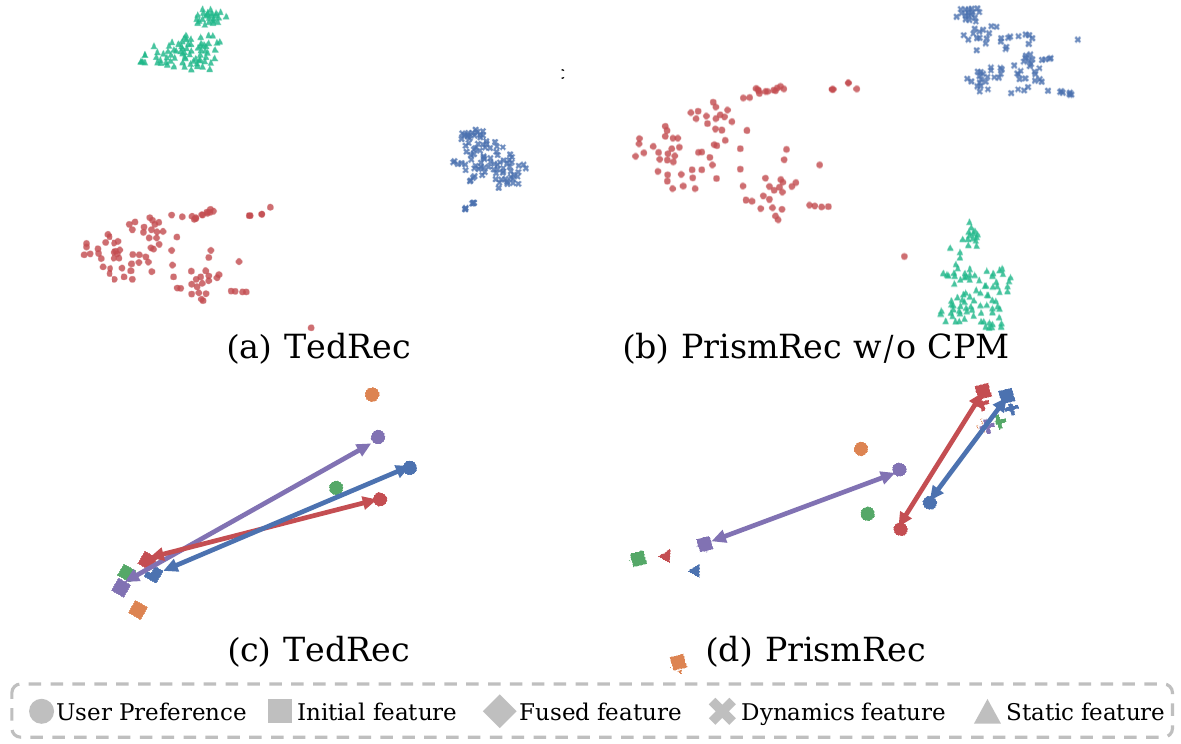}
    \caption{T-SNE visualization of the learned representations on Microlens-Big.}
    \vspace{-0.3cm}
    \label{fig:feature_visualization}
\end{figure}

\subsection{Ablation Study (RQ2)}

To isolate the contribution of each module, we compare our \model with three ablation variants. Specifically, \textit{\model-CPM} substitutes the context-calibrated preference matching with a MLP structure, \textit{\modelwospace-SSF} replaces spectral semantic factorization with the holistic video representation and \textit{\modelwospace-SSF-CPM} removes both. From Figure~\ref{fig:prismrec_ablation}, we can observe that: 
(1) \textit{\modelwospace-SSF} consistently underperforms \modelwospace, indicating that a holistic video representation cannot preserve the complementary static and dynamic cues distributed across frames. Once the two factors are compressed into a unified vector, their distinct roles in preference formation become non-trivial to recover. 
(2) Compared with \modelwospace, \textit{\modelwospace-CPM} also degrades on all datasets, which suggests that a naive MLP characterizes the transition from historical preferences to the next-item representation coarsely. In contrast, CPM refines such transition step by step under content and temporal conditions, thereby producing a preference state better aligned with the target item. 
(3) Notably, the relative contribution of these two modules varies across datasets, and neither one dominates consistently. This is expected given their distinct functions, since SSF operates on the item-side content representation whereas CPM calibrates the user-side preference trajectory, and it further indicates that their benefits are complementary.

\subsection{Universality Analysis (RQ4)} 

To demonstrate the backbone-agnostic applicability of \modelwospace, we instantiate it with SASRec and CL4SRec as sequential backbones\footnote{The experimental results are provided in Appendix.}. Equipping either backbone with SSF or CPM alone improves the average H@10 and N@10 by 16.60\% and 21.01\%, respectively, while integrating both modules achieves the best performance across all settings. This pattern is consistent with the results in Table~\ref{tab:performance_comparison_extended}, indicating that the improvements of \model stem from structured content factorization and context-calibrated transport rather than from a specific sequential encoder.

\begin{figure}[t]
    \centering
    \includegraphics[width=1\linewidth]{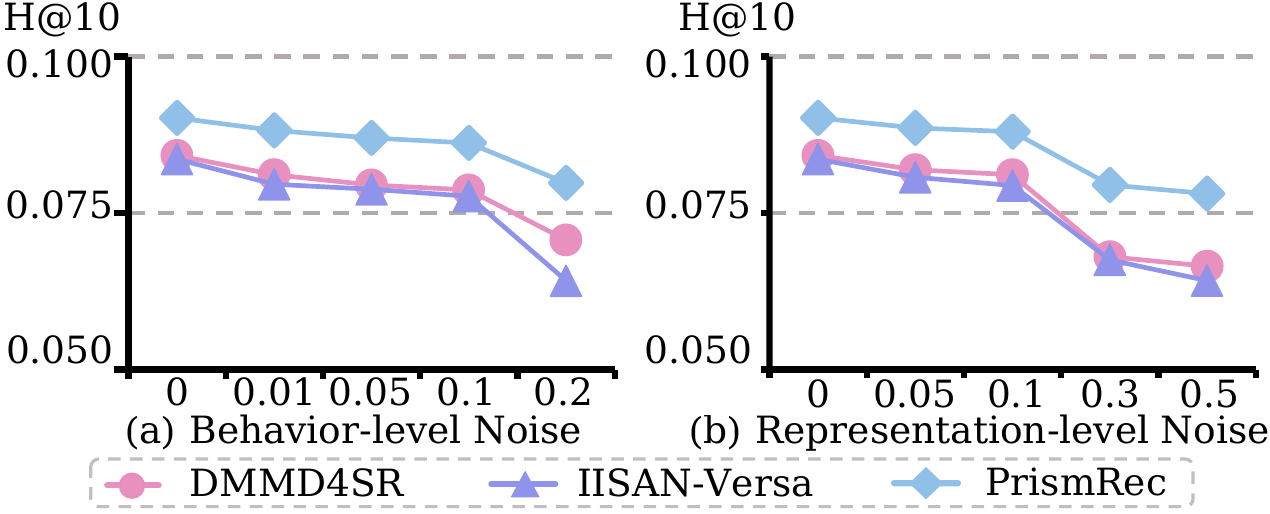}
    \vspace{-0.3cm}
    \caption{Noise robustness of \model and other baselines.}
    \vspace{-0.2cm}
    
    \label{fig:prismrec_noise}
    
\end{figure}

\begin{figure}[t]
    \centering
    \includegraphics[width=1\linewidth]{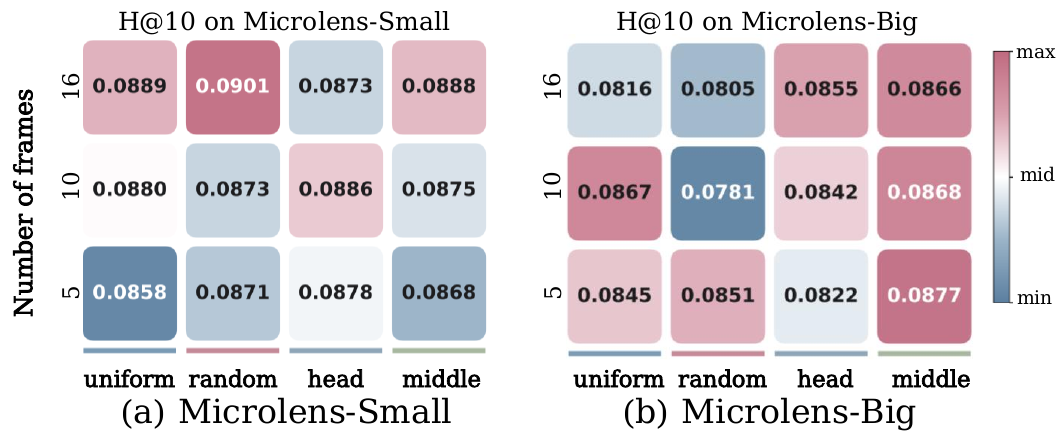}
    \caption{Effects of the frame sampling strategy and budget on H@10. Colors are normalized within each panel.}
    \label{fig:frame_sampling}
\end{figure}

\subsection{Visualization Analysis (RQ3)}

Figure~\ref{fig:feature_visualization} presents a qualitative comparison of the learned representations. Comparing panels (a) and (b), \model without CPM exhibits clearer separation between static and dynamic video factors than TedRec, suggesting that SSF captures complementary temporal characteristics of video content. Comparing panels (c) and (d), the fused representations produced by the complete \model are more closely aligned with the corresponding user preferences. These observations provide qualitative evidence that SSF facilitates static--dynamic factorization, while CPM promotes preference-aware representation alignment.

\subsection{Robustness Analysis (RQ5)}
We evaluate the robustness of \model on Microlens-Small under two noisy settings, where we inject dummy interactions into user behavior sequences at ratios up to 0.2 and add Gaussian perturbations to input representations at scales up to 0.5. From Figure~\ref{fig:prismrec_noise}, we can observe that: (1) \model retains the best H@10 at every noise level, and its margin over DMMD4SR widens from 7.13\% to 12.87\% as the behavior-level ratio grows. Here, SSF derives item-side factors from video content rather than from behavior sparsity, thereby diluting the influence of the injected behaviors. 
(2) Under representation-level noise, the corresponding drops are 13.41\%, 23.21\%, and 21.02\%. CPM refines the preference representation through successive conditioned updates, so that a perturbation carried by any single intermediate representation is attenuated along the transport trajectory. These results jointly verify the robustness of \model against corrupted interactions and noisy representations.

\subsubsection{Parameter Sensitivity.}

Figure~\ref{fig:frame_sampling} examines the joint effects of frame sampling strategy and sampling budget on MicroLens-Small and MicroLens-Big. \model achieves consistently competitive performance across different sampling configurations, while the choice of sampling strategy remains influential due to differences in temporal coverage. Increasing the number of sampled frames within a moderate range provides richer visual evidence and generally improves performance, while sampling additional frames helps sustain these gains by providing more comprehensive visual evidence. These results highlight the importance of balancing representative temporal coverage with sampling density for effective video representation.

\begin{figure}[t]
    \centering
    \includegraphics[width=1\linewidth]{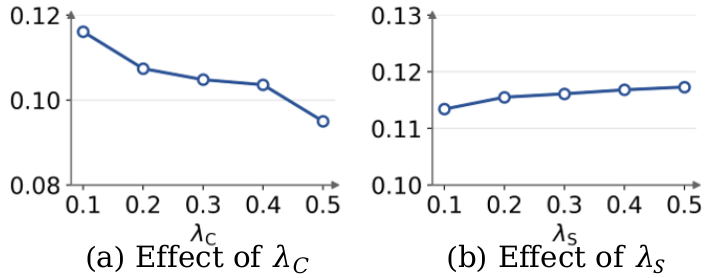}
    \caption{Parameter sensitivity of \model on Microlens-Small in terms of H@20 with respect to $\lambda_C$ and $\lambda_S$.}
    \label{fig:frame_sampling}
\end{figure}

\begin{table}[!t]
    \centering
    \caption{Efficiency comparison results. \#Tra., \#Eval., and \#Mem. denote training and evaluation time per epoch, and the peak GPU memory usage, respectively.}
    \label{tab:complexity}
    \resizebox{\columnwidth}{!}{
    \begin{tabular}{c|c|c|ccc}
        \toprule
        \textbf{Dataset} & \textbf{Algorithms} & \textbf{H@20} & \textbf{\#Tra. (s)} & \textbf{\#Eval. (s)} & \textbf{\#Mem. (MB)} \\
        \midrule
        \multirow{4}{*}{MicroLens-Small} 
        & FMRec & 0.0715 & 5.43 & 15.79 & 4,760 \\
        & IISAN-Versa & \underline{0.1076} & 138.02 & 3.00 & 48,404 \\
        & \textit{PrismRec} & \textbf{0.1177} & 4.43 & 1.84 & 4,276 \\
        \midrule
        \multirow{4}{*}{MicroLens-Big} 
        & FMRec &  \underline{0.1235} & 75.93 & 182.47 & 5,509 \\
        & IISAN-Versa & 0.1146 & 1618.97 & 22.00 & 48,958 \\
        & \textit{PrismRec} & \textbf{0.1289} & 78.54 & 5.70 & 4,691 \\
        \bottomrule
    \end{tabular}
    }
\end{table}

\subsubsection{Complexity Comparison.}
As shown in Table~\ref{tab:complexity}, \modelwospace achieves the best H@20 on both datasets while requiring the least GPU memory and evaluation time. It also provides the fastest training on MicroLens-Small and remains comparable to FMRec on MicroLens-Big, demonstrating a favorable balance between recommendation accuracy and computational efficiency.

\section{Conclusion}
\label{sec:conclusion}
We present \modelwospace, a multimodal flow-matching framework for micro-video sequential recommendation. SSF factorizes frame-level video representations into static- and dynamic-oriented factors, while CPM models preference transitions through a context-calibrated transport process. Experiments on four datasets demonstrate the effectiveness of the proposed method, while robustness and efficiency analyses confirm its practical utility. Future work will explore a more efficient and unified framework for representing heterogeneous information in videos.

\bibliography{aaai2027}

\end{document}